\documentclass[twocolumn,english,prl,aps,floatfix,superscriptaddress,showpacks]{revtex4}
\usepackage[T1]{fontenc}
\usepackage[latin9]{inputenc}
\usepackage{amsmath}
\usepackage{graphicx}
\usepackage{amssymb}
\usepackage{esint}

\makeatletter
\@ifundefined{textcolor}{}
{%
 \definecolor{BLACK}{gray}{0}
 \definecolor{WHITE}{gray}{1}
 \definecolor{RED}{rgb}{1,0,0}
 \definecolor{GREEN}{rgb}{0,1,0}
 \definecolor{BLUE}{rgb}{0,0,1}
 \definecolor{CYAN}{cmyk}{1,0,0,0}
 \definecolor{MAGENTA}{cmyk}{0,1,0,0}
 \definecolor{YELLOW}{cmyk}{0,0,1,0}
 }

\usepackage[dvipdfm]{hyperref}\hypersetup{colorlinks=true,citecolor=blue,linkcolor=blue}

\makeatother

\usepackage{babel}

\begin{document}

\title{Dynamical conductivity at the dirty superconductor-metal quantum
phase transition}

\author{Adrian Del Maestro}

\affiliation{Department of Physics and Astronomy, University of British Columbia,
Vancouver, British Columbia V6T 1Z1, Canada}

\author{Bernd Rosenow}

\affiliation{Institute for Theoretical Physics, University of Leipzig, D-04103
Leipzig, Germany}

\author{José A. Hoyos}

\affiliation{Instituto de Física de São Carlos, Universidade de São Paulo, C.P.
369, São Carlos, São Paulo 13560-970, Brazil}

\author{Thomas Vojta}

\affiliation{Department of Physics, Missouri University of Science and Technology,
Rolla, Missouri 65409, USA}
\begin{abstract}
We study the transport properties of ultrathin disordered nanowires
in the neighborhood of the superconductor-metal quantum phase transition.
To this end we combine numerical calculations with analytical strong-disorder
renormalization group results. The quantum critical conductivity at
zero temperature diverges logarithmically as a function of frequency.
In the metallic phase, it obeys activated scaling associated with
an infinite-randomness quantum critical point. We extend the scaling
theory to higher dimensions and discuss implications for experiments. 
\end{abstract}

\date{\today}


\maketitle
Electrical transport in low-dimensional strongly fluctuating superconductors
has been the subject of intense experimental investigation for almost
half a century \cite{whiskers}. Recently, advances in experimental
techniques have allowed for the fabrication of ultrathin metallic
nanowires having diameters smaller than the bulk superconducting coherence
length, but large enough to include many transverse channels for electronic
conduction. Resistance measurements have shown that the thicker among
these wires exhibit a well-defined phase transition from a resistive
to a superconducting state with decreasing temperature, while thinner
wires appear to remain resistive down to the lowest temperatures measured
\cite{liu-zadorozhny,lau-markovic,boogaard,rogachev-bollinger,chang}.

It has been proposed \cite{swt,drss,drs,drms} that these experiments
may be understood in terms of a superconductor-metal quantum phase
transition (SMT) driven by pair-breaking interactions, possibly due
to random magnetic moments trapped on the wire surface \cite{rogachev-wei}.
A description of this transition is provided by a theory, first proposed
by Feigel'man and Larkin \cite{fl-1}, of a complex Cooper pair order
parameter whose fluctuations are damped by decay into unpaired electrons
\cite{herbut,lsv,spivak-zyuzin,fl-2,galitski}.

As the nanowires are prone to random variations in diameter and because
of the random positions of the pair-breaking moments, quenched disorder
plays an important role. The thermodynamics of the disordered SMT
has been analyzed both analytically \cite{hoyos} and numerically
\cite{drms} in the relevant case of one space dimension. It is governed,
for any nonzero disorder strength, by a non-perturbative infinite-randomness
critical point (IRCP). This IRCP is in the same universality class
as the magnetic quantum critical point of the random transverse-field
Ising chain despite the fact that the two systems have different symmetries:
The clean transverse-field Ising chain can be described by relativistic
free fermions (and, therefore, dynamical exponent $z=1$) whereas
the clean SMT is described by overdamped O(2) fluctuations with $z=2$.
The homology lies in the marginal dynamics of finite size clusters
in both models \cite{vojta-schmalian} which are the famous \emph{rare
regions} of Griffiths-McCoy physics \cite{griffiths}.

Many asymptotically exact results for the random transverse-field
Ising chain \cite{fisher,igloi} apply directly to the SMT via universality.
The IRCP is characterized by activated dynamical scaling: $L_{\Omega}\sim[\ln(\Omega_{0}/\Omega)]^{1/\psi}$.
Here, $\Omega$ is the characteristic energy of the order-parameter
fluctuations on length scale $L_{\Omega}$, $\Omega_{0}$ is a high-energy
reference scale, and $\psi=1/2$ is known as the tunneling exponent.
The exponential length-energy relation implies that the dynamical
exponent $z$ is formally infinite. Moreover, the magnitude of the
order-parameter fluctuations $\mu$ also scales logarithmically with
energy, $\mu_{\Omega}\sim[\ln(\Omega_{0}/\Omega)]^{\phi}$, where
the cluster exponent $\phi=(1+\sqrt{5})/2$ is the golden ratio. Approaching
criticality, the correlation length diverges as $\xi\sim|\delta|^{-\nu}$
where $\nu=2$ and $\delta$ measures the relative distance to the
critical point.

In this Letter, we study experimentally important transport properties
at the pair-breaking SMT of disordered nanowires. We report both analytical
and numerical calculations of the zero-temperature finite-frequency
Aslamazov-Larkin \cite{al} fluctuation corrections to the conductivity
$\sigma\left(\omega\right)$. At criticality, the real part of the
conductivity diverges as $\sigma^{\prime}(\omega)\sim[\ln(\omega_{0}/\omega)]^{1/\psi}$
with vanishing frequency $\omega$ ($\omega_{0}$ is a reference frequency).
Off criticality, it satisfies the unconventional activated scaling
form \begin{equation}
\sigma^{\prime}(\delta,\omega)=\frac{4e^{2}}{h}\left(\ln\frac{\omega_{0}}{\omega}\right)^{1/\psi}\Phi_{\sigma}\left(\delta^{\nu\psi}\ln\frac{\omega_{0}}{\omega}\right),\label{eq:sigma}\end{equation}
 where $\Phi_{\sigma}(x)$ is a universal scaling function. In the
remainder of this Letter, we sketch the derivation of these results
and discuss their experimental implications.


We begin by introducing a one-dimensional continuum model of Cooper
pairs in the presence of Ohmic dissipation and disorder at $T=0$~\cite{swt,drss,drms,hoyos}
\begin{align}
\mathcal{S} & =\int\mathrm{d}x\int\mathrm{d}\tau\left[D(x)\left|\partial_{x}\Psi(x,\tau)\right|^{2}+\alpha(x)\left|\Psi(x,\tau)\right|^{2}\right.\nonumber \\
 & \left.+\frac{u}{2}\left|\Psi(x,\tau)\right|^{4}\right]+\int\mathrm{d}x\int\frac{\mathrm{d}\omega}{2\pi}\gamma(x)\left|\omega\right|\left|\tilde{\Psi}(x,\omega)\right|^{2}.\label{eq:S}\end{align}
 Here, $\tilde{\Psi}(x,\omega)$ is the Fourier transform of $\Psi(x,\tau)$,
a complex superconducting order parameter at position $x$ and imaginary
time $\tau$. We have explicitly included the random spatial dependence
of all coupling constants. Stability and causality constrain $u,\gamma(x)>0$,
and we may choose a gauge where $D(x)>0$. The quantum phase transition
is tuned via $\delta\equiv\alpha-\alpha_{c}$ \cite{drms}.

To proceed, we use a lattice discretization of the continuum action
(\ref{eq:S}) in the limit of a large number of order-parameter components.
This limit has no impact on the character of the critical point \cite{drms,hoyos}
and leads to a quadratic action \begin{equation}
\mathcal{S}_{0}=\sum_{i,j}\int\frac{\mathrm{d}\omega}{2\pi}\tilde{\Psi}_{i}^{\ast}(\omega)(\mathsf{M}_{ij}+|\omega|\delta_{ij})\tilde{\Psi}_{j}^{\phantom{\ast}}(\omega),\label{eq:S0}\end{equation}
 where the coupling matrix $\mathsf{M}_{ij}\equiv(D_{i}/\sqrt{\gamma_{i}\gamma_{j}})\Delta_{ij}^{2}+(r_{i}/\gamma_{i})\delta_{ij}$
(with $\Delta_{ij}$ the discrete nearest neighbor Laplacian) must
be determined self-consistently by solving $r_{i}=\alpha_{i}+(u/\gamma_{i})\left\langle |\Psi_{i}(\tau)|^{2}\right\rangle $
and $\tilde{\Psi}_{i}(\omega)\to\tilde{\Psi}_{i}(\omega)/\sqrt{\gamma_{i}}$
has been rescaled. Note that the full effects of disorder can be realized
while fixing $\gamma_{i}=\gamma$ to be constant \cite{tucker}.

We are now in a position to directly evaluate the dynamical conductivity
via the Kubo formula \cite{kubo} \[
\sigma(\omega)=-\frac{i}{\hbar\omega}\left[\sum_{i,j}\int\mathrm{d}\tau\left\langle J_{i}(\tau)J_{j}(0)\right\rangle \mathrm{e}^{i\omega\tau}-\mathcal{D}\right]_{i\omega\to\omega+i\eta}\]
 where the current is given by $J_{j}(\tau)=(2ie/\gamma\hbar)D_{j}\left[\Psi_{j}^{\ast}(\tau)\Psi_{j+1}^{\phantom{\ast}}(\tau)-\Psi_{j+1}^{\ast}(\tau)\Psi_{j}^{\phantom{\ast}}(\tau)\right]$
with diamagnetic contribution $\mathcal{D}=(8e^{2}/\gamma\hbar)\sum_{i}D_{i}\left\langle |\Psi_{i}(0)|^{2}\right\rangle $.
The Kubo formula can be evaluated by employing the spectral decomposition
of $\mathsf{M}$ in terms of its eigenvector $\mathsf{V}_{ij}$ and
accompanying diagonal eigenvalue $\mathsf{E}_{ij}=\epsilon_{i}\delta_{ij}$
matrices defined by $\sum_{k}\mathsf{M}_{ik}\mathsf{V}_{kj}=\mathsf{V}_{ij}\epsilon_{j}$.
The resulting real part of the conductivity reads \begin{align}
\sigma^{\prime}(\omega) & =\frac{8e^{2}}{h}\sum_{a,b}\sum_{i,j}D_{i}D_{j}\left(\mathsf{V}_{i,a}\mathsf{V}_{j,a}\mathsf{V}_{i+1,b}\mathsf{V}_{j+1,b}\right.\nonumber \\
 & \qquad\left.-\,\mathsf{V}_{i,a}\mathsf{V}_{j+1,a}\mathsf{V}_{i+1,b}\mathsf{V}_{j,b}\right)\mathcal{K}_{a,b}(\omega)\label{eq:sigmaPrime}\end{align}
 where \begin{align}
\mathcal{K}_{a,b}(\omega) & =\left\{ -2\epsilon_{a}(\epsilon_{a}^{2}-\epsilon_{b}^{2}+\omega^{2})\arctan(\omega/\epsilon_{a})\phantom{\frac{a}{b}}\right.\nonumber \\
 & \quad-\,2\epsilon_{b}(-\epsilon_{a}^{2}+\epsilon_{b}^{2}+\omega^{2})\arctan(\omega/\epsilon_{b})\nonumber \\
 & \left.\quad+\omega(\epsilon_{a}^{2}+\epsilon_{b}^{2}+\omega^{2})\ln\left[\frac{(\epsilon_{a}^{2}+\omega^{2})(\epsilon_{b}^{2}+\omega^{2})}{\epsilon_{a}^{2}\epsilon_{b}^{2}}\right]\right\} \nonumber \\
 & \quad\times\left[\epsilon_{a}^{4}+2\epsilon_{a}^{2}(-\epsilon_{b}^{2}+\omega^{2})+(\epsilon_{b}^{2}+\omega^{2})^{2}\right]^{-1}.\label{eq:K}\end{align}

\begin{figure}[t]
\centering{} \includegraphics[clip,height=1\columnwidth,angle=90]{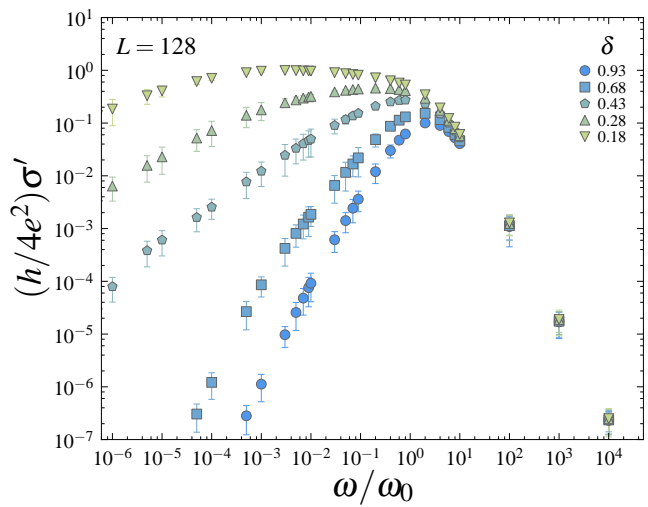}
\caption{\label{fig:conductivity} (Color online) The disorder averaged real
conductivity for chains of 128 sites as a function of frequency measured
in terms of a UV cutoff $\omega_{0}$ for different values of the
distance from the critical point, $\delta$.}

\end{figure}

The validity of the scaling form (\ref{eq:sigma}) can be tested via
a full numerical evaluation of Eqs.~(\ref{eq:sigmaPrime}) and (\ref{eq:K}).
This is possible by exploiting an efficient algorithm for computing
the self-consistent pairing eigenmodes of $\mathcal{S}_{0}$ for large
system sizes \cite{drms}. We have evaluated the conductivity (\ref{eq:sigmaPrime})
for chains of various length with up to $128$ sites averaged over
$3000$ disorder realizations. For clarity, we limit our analysis
to the largest size, $L=128$, as the extrapolation to the thermodynamic
limit $L\to\infty$ is non-trivial due to the crossover between $\xi$
and $L$ for the range of $\delta$ considered here. The results are
displayed in Fig.~\ref{fig:conductivity}.

%


For probe frequencies $\omega$ much larger than the characteristic
fluctuation energy scale $\omega_{0}$ of the chain, we fully saturate
all quantum dynamics and observe a trivial $\sigma^{\prime}(\omega)\sim4e^{2}/(h\omega^{2})$
conductivity. On the other hand for $\omega\ll\omega_{0}$, $\sigma^{\prime}$
appears to be suppressed by a $\delta$-dependent exponent. As we
approach criticality $(\delta\to0)$ the functional form of the average
conductivity is not easily ascertained from Eq.~(\ref{eq:sigmaPrime})
due to the softening of critical modes, but the apparent disappearance
of this exponent is fully consistent with the scaling theory in Eq.~(\ref{eq:sigma}).

\begin{figure}[t]
 \centering \includegraphics[clip,height=1\columnwidth,angle=90]{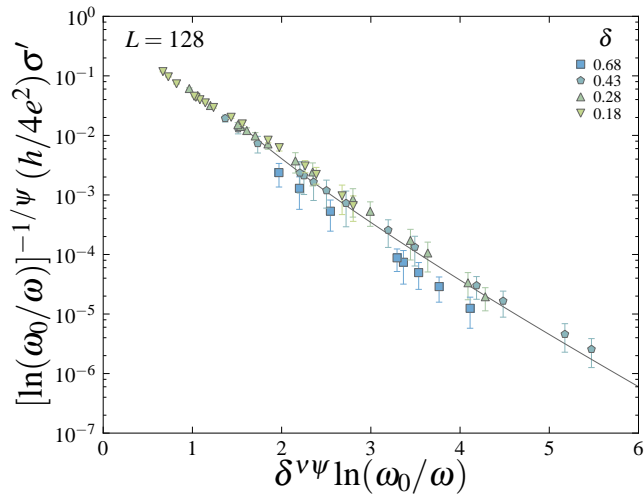}
\caption{\label{fig:conductivityScaling} (Color online) The disorder averaged
real conductivity scaling function $\Phi_{\sigma}$ as a function
of the dimensionless scaling variable $x=\delta^{\nu\psi}\ln(\omega_{0}/\omega)$
for different values of $\delta$ as criticality is approached from
above. The line is a guide to the eye showing the probable functional
form of $\Phi_{\sigma}$ (see text).}

\end{figure}

The predictions of Eq.~(\ref{eq:sigma}) can be more thoroughly confirmed
by searching for consistent scaling of the data shown in Fig.~\ref{fig:conductivity}
after dividing by $[\ln(\omega_{0}/\omega)]^{1/\psi}$ and re-plotting
as a function of the dimensionless scaling variable $x\equiv\delta^{\nu\psi}\ln(\omega_{0}/\omega)$.
We find excellent data collapse over five orders of magnitude as shown
in Fig.~\ref{fig:conductivityScaling}. 

From the scaled data, we extract the universal prefactor of the conductivity
at the critical point, $\Phi_{\sigma}(0)=0.70(4)$. Furthermore, an
empirical analysis of the numerical scaling function in Fig.~\ref{fig:conductivityScaling}
suggests that $\Phi_{\sigma}(x\to\infty)\sim x^{-\nu}\mathrm{e}^{-\mathcal{A}x}$
where $\mathcal{A}\sim\mathrm{O}(1)$. The relevant limits of $\Phi(x)$
can also be inferred by appealing to the naive scaling prediction
that in $d=1$ the conductivity should be equal to $4e^{2}/h$ multiplied
by a length. Activated dynamical scaling dictates that at criticality,
lengths scale like $[\ln(\omega_{0}/\omega)]^{1/\psi}$ whereas in
the Griffiths phase the relevant length scale is the correlation length
$\xi\sim|\delta|^{-\nu}$, fixing the $x$-dependent power in front
of the exponential in $\Phi_{\sigma}(x\to\infty)$ in order to cancel
the logarithmic prefactor in Eq.~(\ref{eq:sigma}).

Let us now turn to an analytical derivation of the dynamical conductivity.
Near an IRCP, the conductivity will be dominated by large rare regions
which are \emph{locally} in the superconducting phase, i.e., by small
clusters with exceptionally strong links $D$ and typically small
gaps $r$. In the low frequency limit, the effective links and gaps
of these clusters can be quantified by a renormalization group analysis
\cite{hoyos} that we invoke later. For now, we approximate each dominant
cluster as a single two-site system with $r_{1,2}$ being the effective
local gaps and $D$ the effective link strength. In such a simple
model, the conductivity Eq.~(\ref{eq:sigmaPrime}) can be evaluated
exactly, resulting in \begin{equation}
\sigma_{\mathrm{2site}}^{\prime}=\frac{8e^{2}D^{2}\ell^{2}}{\gamma^{2}\omega}\mathcal{K}_{12}(\omega),\label{eq:sigma2}\end{equation}
 where the eigenvalues of the $2\times2$ coupling matrix are $2\epsilon_{1,2}=[D/\gamma+(r_{1}+r_{2})/\gamma\pm\sqrt{(D/\gamma)^{2}+(r_{1}-r_{2})^{2}/\gamma^{2}}]$
and we have introduced $\ell$ as the \emph{length} of the link connecting
the two sites. Due to the presence of the factor $D^{2}$ in Eq.~(\ref{eq:sigma2}),
the average over the contributions of all two-site clusters will be
dominated by those with an anomalously large links $D\gg r_{1,2}$.
Hence, we may evaluate the conductivity by averaging \begin{equation}
\sigma^{\prime}\left(\Omega\right)=n_{\Omega}\int_{0}^{\Omega}{\rm d}D\int_{0}^{\infty}{\rm d}\ell\, P\left(D,\ell\right)\int_{0}^{D}{\rm d}rR\left(r\right)\sigma_{\mathrm{2site}}^{\prime},\label{eq:sigmaMean}\end{equation}
 where $\Omega=\gamma\omega$ is the energy scale at which effective
clusters with gap $r$ connected by links of magnitude $D$ and length
$\ell$ appear with probability density $R\left(r\right)P\left(D,\ell\right)$,
and $n_{\Omega}$ is the density of such clusters. The final step
consists of using the asymptotic value of $\mathcal{K}_{12}(\omega)\sim\omega^{-1}\ln[\Omega/r]$,
in the strong disorder limit and $\Omega,D\gg r$ where $r=(r_{1}+r_{2})/2$.
Using the values of $P(D,\ell)$, $R(r)$ and $n_{\Omega}$ at criticality
\cite{fisher,igloi}: $P(y)=\int{\rm d}\ell P(y,\ell)=e^{-y}$, with
$y=\ln(\Omega/D)/\ln(\Omega_{0}/\Omega)$, $R(y)=P(y)$ as well as
the relation $n_{\Omega}^{-1}\sim\left\langle \ell\right\rangle \sim[\ln(\Omega_{0}/\Omega)]^{1/\psi}$
between density and average separation of clusters, we arrive at $\sigma^{\prime}\sim n_{\Omega}\left\langle (D/\Omega)^{2}\ell^{2}\right\rangle \left\langle \ln\left(\Omega/r\right)\right\rangle \sim[\ln(\Omega_{0}/\Omega)]^{1/\psi}$,
where $1/\psi=2$ recovering Eq.~(\ref{eq:sigma}) at criticality
where $\delta^{\nu\psi}\ln(\Omega_{0}/\Omega)\to0$. Correlations
between the links $D$ and their lengths $\ell$ have no effect on
the leading logarithmic divergence of this result at criticality.
The analysis of Eq.~(\ref{eq:sigmaMean}) in the metallic and superconducting
Griffiths phases crucially depends on the careful treatment of the
correlations between $\ell$ and $D$. The resulting expressions are
quite involved and will be discussed elsewhere \cite{griffTransport}.

We now discuss a sub-leading correction to the scaling of $\sigma^{\prime}\left(\omega\right)$.
In Eq.~(\ref{eq:sigmaMean}), the dissipative $z=2$ dynamics causes
the relation between energy $\Omega$ and the measured frequency $\omega$
to have a logarithmic correction \cite{hoyos} $\Omega=\gamma_{0}\mu_{\Omega}\omega$,
where $\gamma_{0}$ is the bare dissipative coupling and $\mu_{\Omega}$
is the mean value of superconducting order-parameter fluctuations.
At criticality $\mu_{\Omega}\sim[\ln(\Omega_{0}/\Omega)]^{\phi}$
and $\ln(\Omega_{0}/\Omega)=\ln(\omega_{0}/\omega)$ up to log(log)
corrections. In the metallic Griffiths phase, where presumably any
real experiments on metallic nanowires would take place, $\mu_{\Omega}\sim\delta^{\nu\psi(1-\phi)}\ln(\Omega_{0}/\Omega)$
and the logarithms of energy and frequency are no longer simply equivalent.
However, the exact value of $\mu_{\Omega}$ can be obtained from the
imaginary part of the dynamical order-parameter susceptibility \cite{hoyos,drms},
and its inclusion leads to quantitatively better data collapse as
we extend the scaling theory of Eq.~(\ref{eq:sigma}) deeper into
the Griffiths regime \cite{griffTransport}.

In order to place the appearance of the link length $\ell$ in Eq.~(\ref{eq:sigma2})
and the average of Eq.~(\ref{eq:sigmaMean}) on firmer footing, we
invoke the real space renormalization group technique of Refs.~\cite{hoyos,dmh},
providing direct access to the renormalization of the current operator.
Starting with a chain described by the effective action of Eq.~(\ref{eq:S0}),
we proceed by searching for the largest local coupling of the chain,
$\Omega=\max\left\{ r_{i},D_{i}\right\} $. Suppose (i) $\Omega=r_{2}$,
site $2$ is then strongly fluctuating and can be integrated out of
the the system leading to an effective coupling $\tilde{D}=D_{1}D_{2}/r_{2}$
between sites $1$ and $3$. Their relative distance is given by $\tilde{\ell}=\ell_{1}+\ell_{2}$,
and the total current through clusters $1-3$ is $\ell_{1}J_{1}+\ell_{2}J_{2}=\tilde{\ell}\tilde{J}$,
where $\tilde{J}=(2ie/\gamma\hbar)\tilde{D}[\Psi_{1}^{\ast}\Psi_{3}^{\phantom{\ast}}-\Psi_{3}^{\ast}\Psi_{1}^{\phantom{\ast}}]$.
If on the other hand (ii) $\Omega=D_{2}$, sites $2$ and $3$ are
strongly coupled forming an effective cluster $\tilde{2}$ in which
the effective gap is $\tilde{r}_{2}=r_{2}r_{3}/D_{2}$. The total
current is then $\ell_{1}J_{1}+\ell_{2}J_{2}+\ell_{3}J_{3}=\tilde{\ell}_{1}\tilde{J}_{1}+\tilde{\ell}_{2}\tilde{J}_{2}$,
where $\tilde{\ell}_{1}=\ell_{1}+\frac{1}{2}\ell_{2}$, $\tilde{\ell}_{2}=\frac{1}{2}\ell_{2}+\ell_{3}$,
$\tilde{J}_{1}=(2ie/\gamma\hbar)D_{1}[\Psi_{1}^{\ast}\Psi_{\tilde{2}}^{\phantom{\ast}}-\Psi_{\tilde{2}}^{\ast}\Psi_{1}^{\phantom{\ast}}]$
and $\tilde{J}_{2}=(2ie/\gamma\hbar)D_{3}[\Psi_{\tilde{2}}^{\ast}\Psi_{4}^{\phantom{\ast}}-\Psi_{4}^{\ast}\Psi_{\tilde{2}}^{\phantom{\ast}}]$.
After process (i) or (ii), the energy scale $\Omega$ is lowered and
the disorder of the effective system is increased \cite{fisher}.
Iterating this procedure leads to the probability distribution of
gaps and links connecting the effective clusters as a function of
energy \cite{hoyos} providing formal justification of Eq.~(\ref{eq:sigmaMean}).

We now compare our results with the transport properties of other
systems governed by infinite randomness. To the best of our knowledge,
only one other such system has been studied: spin conductivity in
the dimerized antiferromagnetic spin-1/2 chain \cite{dmh}. At criticality,
the spin conductivity also diverges logarithmically, but with a weaker
power than here, and is found to obey the scaling form $\sigma_{\mathrm{spin}}(\omega)\sim\ln(\omega_{0}/\omega)\Phi_{\mathrm{spin}}\left[\delta\ln(\omega_{0}/\omega)\right]$,
where $\Phi_{\mathrm{spin}}(x\to0)\approx7/180$ and $\Phi_{\mathrm{spin}}(x\to\infty)\sim xe^{-2x}$.
Although the thermodynamics of our system and the spin chain are funneled
into the same universality class by the disorder, their transport
properties are \emph{not} universal because their underlying dynamics
are different.

Finally, we highlight that the methods discussed here should also
apply in higher dimensions. Specifically, at criticality, the system
will also be governed by an infinite-randomness critical point \cite{hoyos,igloi}
but with different exponents $\psi$, $\nu$, and $\phi$. The dynamical
conductivity will likewise be dominated by rare and locally superconducting
strongly coupled regions. Evaluating Eq.~(\ref{eq:sigmaMean}) immediately
leads to $\sigma^{\prime}(\omega)\sim[\ln(\omega_{0}/\omega)]^{(2-d)/\psi}$,
since the spatial dimension enters explicitly only via the density
of clusters $n_{\Omega}\sim L_{\Omega}^{-d}$, and the average conductivity
of a two-site cluster $\langle\sigma_{{\rm 2site}}^{\prime}\rangle\sim[\ln(\omega_{0}/\omega)]^{2/\psi}$.
As an immediate consequence, in the limit $\omega\to0$, the critical
conductivity vanishes for $d>2$, and becomes constant at $d=2$.

In conclusion, we have presented the numerically evaluated dynamical
conductivity $\sigma^{\prime}(\omega)$ for a model believed to describe
the physics of disordered nanowires close to a superconductor-metal
quantum phase transition and placed it in an analytical framework
computed via the strong disorder renormalization group. We have shown
that $\sigma^{\prime}(\omega)$ diverges logarithmically as $\left|\ln\omega\right|^{2}$
at criticality and obeys scaling in the metallic (Griffiths) phase
with asymptotics dictated by naive dimensional analysis of physical
quantities. Our results may be directly applicable to experimental
transport measurements on thin nanowires which remain metallic as
$T\to0$, such as those reported in Ref.~\cite{chang}. By studying
wires of varying thickness at low temperatures, it may be possible
to reach the critical regime where the logarithmic divergence of the
fluctuation correction to the conductivity for $\omega>T$ could be
directly observed. In $d=2$, we expect $\sigma^{\prime}$ to be frequency-independent,
consistent with experiments in disordered thin superconducting films
(see, e.g.\ Ref.\ \cite{armitage}). Finally, by presenting solid
predictions arising from an effective action of strongly repulsive
dissipative Cooperons, we lay open an avenue for the experimental
investigation of the efficacy of such models when applied to dirty
low dimensional superconductors.

We thank G.~Refael and S.~Sachdev for stimulating and insightful
discussions. We acknowledge the hospitality of the Max-Planck-Institute
for Solid State Research in Stuttgart and the Aspen Center for Physics.
Financial support was provided by FAPESP under grant No. 2010/03749-4,
CNPq under grant No. 302301/2009-7, Research Corporation and NSF under
grant nos. DMR-0339147 and DMR-0906566.

\end{document}